\documentclass[12pt]{article}

\newif\iffigs\figstrue

\usepackage{epsfig,latexsym}
\usepackage{amsmath}
\usepackage{color}
\usepackage{graphicx}
\usepackage{verbatim}
\usepackage{mathrsfs}
\usepackage{amssymb}

\DeclareMathAlphabet{\mathpzc}{OT1}{pzc}{m}{it}

 \csname
@addtoreset\endcsname{equation}{section}

\def\gz0{\gamma^{0}}

\def\beq{\begin{equation}}
\newcommand{\eeq}[1]{\label{#1}\end{equation}}
\def\bea{\begin{eqnarray}}
\newcommand{\eea}[1]{\label{#1}\end{eqnarray}}
\def\ba{\begin{array}}
\def\ea{\end{array}}
\def\bec{\begin{center}}
\def\ec{\end{center}}
\def\ba{\begin{align}}
\def\ena{\end{align}}

\def\12{\frac{1}{2}}

\usepackage{slashed}

\usepackage{float}

\usepackage{booktabs}% for top-,mid- & bottomrule
\usepackage{caption}

\usepackage[mathscr]{euscript}

\usepackage{color}
\usepackage{graphicx}
\usepackage{epsf}
\usepackage{graphicx,epsfig}
\usepackage{amsmath}

\usepackage{multirow}

\usepackage{amsmath, amsthm, amssymb}

\usepackage{epsfig}
\usepackage{cite}
\usepackage{color,colordvi}

\newcounter{hran}

\makeatletter
\renewcommand\section{\@startsection {section}{1}{\z@}%
                               {-3.5ex \@plus -1ex \@minus -.2ex}%
                               {2.3ex \@plus.2ex}%
                               {\normalfont\large\bfseries}}
\makeatother

\vspace{0.5cm}

\newcommand{\bi}{\begin{itemize}}
\newcommand{\ei}{\end{itemize}}

\begin{document}
\thispagestyle{empty}
\begin{flushright}
CERN-PH-TH/2015-214\\
{\today}
\end{flushright}

\vspace{15pt}

\begin{center}

%%%%%%%%%%%%%%%%%%%%%%%%%%%%%%%%%%%%%%%%%%%%%%%%%%%%%%%%%%%%%%%%%%%%

{\Large\sc Supersymmetry and Inflation}\\

%%%%%%%%%%%%%%%%%%%%%%%%%%%%%%%%%%%%%%%%%%%%%%%%%%%%%%%%%%%%%%%%%%%%

\vspace{35pt}
{\sc S.~Ferrara${}^{\; a,b,c}$ and A.~Sagnotti${}^{\; d}$}\\[15pt]

{${}^a$\sl\small Th-Ph Department, CERN\\
CH - 1211 Geneva 23, SWITZERLAND \\ }
\vspace{6pt}

{${}^b$\sl\small INFN - Laboratori Nazionali di Frascati \\
Via Enrico Fermi 40, I-00044 Frascati, ITALY}\vspace{6pt}

{${}^c$\sl\small Department of Physics and Astronomy, \\ U.C.L.A., Los Angeles CA 90095-1547,
USA}\vspace{6pt}

{${}^d$\sl\small Scuola Normale Superiore and INFN,\\
Piazza dei Cavalieri 7\
I-56126 Pisa, Italy}

%%%%%%%%%%%%%%%%%%%%%%%%%%%%%%%%%%%%%%%%%%%%%%%
\vspace{30pt}

%%%%%%%%%%%%%%%%%%%%%%%%%%%%%%%%%%%%%%%%%%%%%%%
{\sc\large Abstract} \end{center}
\noindent Theories with elementary scalar degrees of freedom seem nowadays required for simple descriptions of the Standard Model and of the Early Universe. It is then natural to embed theories of inflation in supergravity, also in view of their possible ultraviolet completion in String Theory. After some general remarks on inflation in supergravity, we describe examples of minimal inflaton dynamics which are compatible with recent observations, including higher--curvature ones inspired by the Starobinsky model. We also discuss different scenarios for supersymmetry breaking during and after inflation, which include a revived role for non--linear realizations. In this spirit, we conclude with a discussion of the link, in four dimensions, between ``brane supersymmetry breaking'' and the super--Higgs effect in supergravity.
\baselineskip=14pt

\vskip 36pt
{\sl
\noindent \small Based in part on the Plenary and Parallel Session talks given by S.~F. at the ``Fourteenth Marcel Grossmann Meeting - MG14'', Rome, July 12-18 2015, on the talk given by S.~F. at ``The String Theory Universe'', 21st European String Workshop, Leuven, September 7-11, 2015, and on the plenary talk given by A.~S. at ``Planck 2015'', Ioannina, May 25-29 2015.}
%%%%%%%%%%%%%%%%%%%%%%%%%%%%%%%%%%%%%%%%%%%%%%%
\vfill
\noindent
\baselineskip=20pt
\setcounter{page}{1}

\newpage

\tableofcontents
\newpage
%%%%%%%%%%%%%%%%%%%%%%%%%%%%%%%%%%%%%%%%%%%%%%%%%%%%%%%
\section{Introduction}
%%%%%%%%%%%%%%%%%%%%%%%%%%%%%%%%%%%%%%%%%%%%%%%%%%%%%%%

In this report we describe a number of approaches to inflaton dynamics based on supergravity \cite{supergravity}, the combination of supersymmetry with General Relativity (GR). Inflationary Cosmology \cite{inflation} affords a convenient description in terms of a single real scalar field, the inflaton, evolving in a Friedmann, Lemaitre, Robertson, Walker (FLRW) geometry. A fundamental scalar field describing a Higgs particle was also recently discovered at LHC, confirming the interpretation of the Standard Model as a spontaneously broken phase (BEH mechanism) of a non-abelian Yang-Mills theory \cite{beh}.

There is thus some evidence that Nature is inclined to favor, both in Cosmology and in Particle Physics, theories which use scalar degrees of freedom, even in diverse ranges of energy scales. Interestingly, there is even a cosmological model where the two degrees of freedom, inflaton and Higgs, are identified, the Higgs inflation model \cite{bs}, where a non-minimal coupling $h^2\, R$ of the Higgs field $h$ to gravity is introduced.

Another model based on an $R + R^2$ extension of GR is the Starobinsky model \cite{inflation}. It is conformally equivalent to GR coupled to a scalar field, the scalaron \cite{whitt}, with a specific form of the scalar potential that drives inflation:
\beq
V \, =\, V_0 \left(1 \ - \
e^{\,-\,\sqrt{\frac{2}{3}}\ \varphi} \right)^2 \qquad
V_0 \sim 10^{-9} {\rm \ in \ Planck \ units} \ .
\eeq{1}
These two models (and also a more general class) give identical predictions for the slow-roll parameters (see table I for their definitions).

An interesting modification of the Starobinsky potential, suggested by its embedding in $R + R^2$ supergravity \cite{fklp}, involves and $\alpha$--deformed potential
\beq
V_\alpha \, =\, V_0
\left(1 \ - \ e^{\,-\,\sqrt{\frac{2}{3\,\alpha}}\
\varphi}\right)^2 \ .
\eeq{5}
It gives the same result for $n_s$ as \eqref{1}
\beq
n_s \ = \ 1 \ - \ \frac{2}{N} \ ,
\eeq{501}
but now \cite{fklp}
\beq
r \ = \ \frac{12\,\alpha}{N^2} \ .
\eeq{6}
This family of models provides an interpolation between the Starobinsky model (for $\alpha$=1) and Linde's chaotic inflation model \cite{inflation} with quadratic potential, which is recovered as $\alpha \to \infty$:
\beq
V_\alpha \ \ \rightarrow \ \ m^2\,
\varphi^2 \quad \left(V_0/\alpha \
{\rm fixed}\right) \qquad \left( {\rm same\ } n_s {\rm \ \ but \ \ } r =\frac{8}{N} \right) \ .
\eeq{7}
The 2015 data analysis from Planck \cite{planck15} favors the Starobinsky model \cite{inflation} with $n_s \simeq 0.97$, $r < 0.11$.

The expression \eqref{7} for $\alpha$--attractor potentials $V_\alpha$ can be further generalized \cite{klr} introducing an arbitrary monotonically increasing function $f\left( \tanh \frac{\varphi}{\sqrt{6\,\alpha}}\right)$,
so that
\beq
V_\alpha(\varphi) \ = \
f^2\left( \tanh \frac{\varphi}{\sqrt{6\,\alpha}}\right) \ , \
\varphi \to \infty \ : \ f\left( \tanh \frac{\varphi}{\sqrt{6\,\alpha}}\right) \to 1 \ - e^{-\sqrt{\frac{2}{3\,\alpha}}\, \varphi}\, + \, \ldots \ .
\eeq{8}

In Sections 2 and 3 we describe several extensions of these ``single field'' inflationary models in the framework of $N=1$ supergravity, with special attention to the problem of embedding the inflaton $\phi$ in a supermultiplet and to the role of the remaining superpartners.

In this context, inflationary models are embedded in a general supergravity theory coupled to matter in FLRW geometries. Under the assumption that no additional supersymmetry ($N \geq 2$) is restored in the Early Universe, the most general $N=1$ extension of GR is obtained coupling the graviton multiplet $(2,3/2)$ to a certain number of chiral multiplets $(1/2,0,0)$, whose complex scalar fields are denoted by $z_i$, $i=1...N_s/2$, and to (gauge) vector multiplets (1,1/2), whose vector fields are denoted by $A_\mu^\Lambda$ $(\Lambda=1,..,N_V)$.
\begin{table}
\centering
\begin{tabular}{ccc}
\hline\hline
 \\
$ \epsilon \,=\, \frac{{M_{P}}^2}{2} \left( \frac{V^\prime}{V}
\right)^2$ & & $\eta \,=\, M_{P}^2 \left( \frac{V^{\prime\prime}}{V}\right)$ \\[-3pt] \\[-3pt]
%\hline \\
& $N \, \simeq \, \frac{1}{M_P^2} \ \int_{\varphi_{\rm end}}^\varphi
\frac{V}{V^\prime} \ d \varphi$ &
%\hline
  \\[-3pt] \\[-3pt]
% \hline \\
$ n_s \ \simeq \  1\ - \ 6\, \epsilon \ +
\ 2\, \eta$ & &
$r \ \simeq \ 16\, \epsilon$ \\[-3pt]
\\[-3pt]
\hline\hline
\end{tabular}
%}
\caption{\small Slow--Roll Inflationary Parameters}
\label{table_1}
\end{table}
These multiplets can acquire supersymmetric masses, and in this case the massive vector multiplet becomes $(1,2(1/2),0)$, eating a chiral multiplet in the supersymmetric version of the BEH mechanism.

For Cosmology, the most relevant part of the Lagrangian \cite{cfgvp,bagger} is the sector which contains the scalar fields coupled to the Einstein-Hilbert action
\beq
e^{-1}\,{\cal L} \ = \ - \ R\ - \ \partial_i \, \partial_{\bar{j}}\,K \, D_\mu z^i\, D_\nu \overline{z}^{\bar{j}} \, g^{\mu\nu} \ - \ V(z,\overline{z}) \ + \ \ldots \ .
\eeq{9}
where $e$ is the determinant of the vierbein, $K$ is the K\"ahler potential of the manifold of scalar fields, and the ``dots'' stand for fermionic terms and gauge interactions.

The scalar covariant derivative is $D_\mu z^i \ = \ \partial_\mu z^i \ + \delta_\Lambda z^i \, A_\mu^\Lambda$, where the $\delta_\Lambda z^i$ are Killing vectors. This term allows to write massive vector multiplets \emph{\`a la} Stueckelberg.
The scalar potential is
\bea
&&V(z^i,\overline{z}^{\bar{i}}) \ = \ e^G \left[ G_i \, G_{\bar{j}} \left( G^{-1}\right)^{i\,\bar{j}} \ - \ 3 \right] \ + \ \frac{1}{2}  \left( Re f_{\Lambda\Sigma}\right)^{-1} D_\Lambda\, D_\Sigma \ , \nonumber \\
&& G \ = \ K \ + \ \log \left| W \right|^2 \ , \ \ W(z^i){\rm \: \ superpotential} \ , \ \ G_{i\bar{j}} = \partial_i \,\partial_{\bar{j}} \, K \ .
\eea{10}
The first and third non-negative terms are referred to as \emph{F} and \emph{D}--term contributions: they explain the possibility of having unbroken supersymmetry in Anti-de Sitter space.

The potential can be recast in the more compact form
\beq
V(z^i,\overline{z}^{\bar{j}}) \ = \ F_i \, F^i \ + \ D_\Lambda\, D^\Lambda \ - \
3\, \left| W \right|^2 \, e^K \ ,
\eeq{11}
with
\beq
F_i \ = \ e^\frac{K}{2} \left( W\, K_{,i}  \ +  \ W_{,i} \right) \ ,
\quad D_\Lambda = G_{,i}\ \delta_\Lambda z^i \ .
\eeq{12}
In a given phase (which could be the inflationary phase of the exit from inflation) broken supersymmetry requires that the equations
\beq
F_i \ = \ D_\Lambda \ = \ 0 \ ,
\qquad \ V \ = \ - \ 3\, |W|^2 \, e^K
\eeq{13}
have no solutions. On the other hand, if a solution exists the possible vacua correspond to Minkowski or AdS phases depending on whether $W$ vanishes or not.

In general, in phases with broken supersymmetry one can have AdS, dS or Minkowski vacua.  Therefore, one can accommodate both the inflationary phase (dS) and the Particle Physics phase (Minkowski). However, it is not trivial to construct corresponding models, since the two scales are very different if supersymmetry is at least partly related to the Hierarchy problem.

In view of the negative contribution $-\ 3\, e^G$ present in the scalar potential, it may seem impossible (or at least not natural) to retrieve a scalar potential exhibiting a de Sitter phase for large values of a scalar field to be identified with the inflaton. The supersymmetric versions of the $R+R^2$ (Starobinsky) model show two solutions for this puzzle: the theory can have (with $F$-terms) a no-scale structure \cite{noscale}, which makes the potential positive along the inflationary trajectory \cite{cecotti}, or alternatively the potential can be a pure $D$-term, which is clearly positive \cite{cfps}. These types of models and their generalizations contain two chiral superfields $(T,S)$ \cite{eno,kl}, as in the old minimal version of $R+R^2$ supergravity \cite{cecotti}, or one massive vector multiplet \cite{fklp,fkr}, as in the new minimal version. Unbroken supersymmetry is recovered in a Minkowski vacuum at the end of inflation.

Another option to attain positive definite potentials, discussed in Section \ref{sec:sgoldstinoless}, is to have sgoldstino-less models where supersymmetry is non--linearly realized. In this class of models the multiplet S, which does not contain the inflaton ($T$ multiplet), is replaced by a nilpotent superfield ($S^2=0$) \cite{quadratic}: this eliminates the sgoldstino scalar from the theory but its $F$-component still drives inflation or at least participates in it. This mechanism was first considered in a different context, where the inflaton resided in a constrained superfield \cite{ag_cosmo}, and was applied
to the Starobinsky model in \cite{adfs}, replacing the S field by a Volkov-Akulov \cite{va} nilpotent superfield, and then to general $F$-term induced inflationary models \cite{fkl}. The result will be referred to, in the following, as the V-A-S model. Recently progress was made \cite{kl14,dz} in the embedding of two different supersymmetry breaking scales in the inflationary potential, in the framework of nilpotent inflation \cite{adfs}. Some of the material presented here may also be found in \cite{fs_vara}.

\begin{table}
\centering
$$ {\cal L} \ \sim \ \left. - \Phi \, S_0\, \overline{S}_0  \right|_D \ + \ \left. \left(W\, S_0^3 + {\rm h.c.} \right) \right|_F \ , \quad \Phi = \exp\left(\,-\, \frac{K}{3}\right)$$
%\vskip 18pt
%\scalebox{0.8}{
\begin{tabular}{ccc}
\hline\hline
Higher Curvature & & Standard Supergravity \\
\hline\hline
 \\
%\hline
 $\Phi_H = 1 - h\left(\frac{{\cal R}}{S_0},\frac{\overline{{\cal R}}}{\overline{S}_0}\right)$ & &$\Phi_S = 1 + T + \overline{T} - h(S,\overline{S})$ \\[-4pt] \\[-4pt]
 $W_H = W\left(\frac{{\cal R}}{S_0}\right)$ & & $W_S = T S + W(S)$ \\[-4pt] \\[-4pt]
 \hline \\[-4pt]
 $\Phi_H = 1 $ & & $\Phi_S = 1 + T + \overline{T} $ \\[-4pt] \\[-4pt]
 $W_H = W\left(\frac{{\cal R}}{S_0}\right)$ & & $W_S =  \left. -S\, {W^\prime}(S) + W(S)\right|_{T=-{W^\prime}(S)}$ \\[-4pt] \\[-4pt]
 \hline \\[-4pt]
 $\Phi_H = - \alpha\, \frac{{\cal R}}{S_0}\ \frac{\overline{{\cal R}}}{\overline{S}_0} $ & & $\Phi_S = T + \overline{T} - \alpha S \, \overline{S} $ \\[-4pt] \\[-4pt]
 $W_H = - \beta\, \frac{{\cal R}^3}{S_0^3}$ & & $W_S =  T S - \beta \, S^3$ \\[-4pt]
\\[-4pt]
\hline\hline
\end{tabular}
%}
\caption{\small Old--Minimal Dualities}
\label{table_2}
\end{table}

\section{Minimal models for Inflation and Supersymmetry Breaking}

In this Section we describe some models of inflation that rest on $N=1$ Supergravity. We begin in Section \ref{sec:sgoldstino_infl} with a brief overview of models containing a single chiral multiplet that are based on the old--minimal formulation \cite{oldminimal}, and then in Section \ref{sec:dterm} we discuss models inspired by the new--minimal formulation \cite{newminimal}. In Section \ref{sec:othermodels} we describe the extension of $\alpha$--attractors allowing for $F$--term breakings and models with flat K\"ahler geometry with a shift symmetry that avoids the $\eta$--problem \cite{shift_eta}. In Section \ref{sec:sgoldstinoless} we discuss nilpotent inflation, where the sgoldstino superfield is subject to a quadratic constraint, so that supergravity is coupled to a Volkov--Akulov \cite{va} non--linear multiplet.

\subsection{Sgoldstino Inflation} \label{sec:sgoldstino_infl}
This class includes models in which the inflaton is identified with the sgoldstino and only one chiral multiplet $T$ is used (Table \ref{table_2}). However, the $f({\cal R})$ supergravity models \cite{ketov,fkvp} yield potentials that either have no plateau or,  when they do, lead to AdS rather than dS phases \cite{fkp_13}. This reflects the no-go theorem of \cite{eno}.

A way out of this situation was recently found with “$\alpha$-scale Supergravity” \cite{roestsca}: adding two superpotentials W+ + W-
which separately give a flat potential along the inflaton (Re$T$) direction gives rise to a de Sitter plateau for large Re$T$. The problem with these models is that the inflaton trajectory is unstable in the ImT direction, but only for small inflaton field: modifications to the superpotential are advocated to generate an inflationary potential.
For single-field models and related problems, see also \cite{achucarro,ketovter,linde_alpha}.
\begin{table}
\centering
%\vskip 18pt
%\scalebox{0.8}{
\begin{tabular}{ccc}
\hline\hline
Higher Curvature & & Standard Supergravity \\
\hline\hline
 \\
%\hline
 $\left.L \log \left(\frac{L}{S_0\overline{S}_0}\right) \right|_D$ & & $\exp\left(\,-\, \frac{K}{3}\right) = - \, U \, \exp U $ \\[-4pt] \\[-4pt]
 $\left. W^\alpha\left(\frac{L}{S_0\overline{S}_0}\right) \, W_\alpha\left(\frac{L}{S_0\overline{S}_0}\right)\right|_F$ & & $W^\alpha(U)\, W_\alpha(U)$ \\[-4pt] \\[-4pt]
 \hline \\[-4pt]
$ $ & & $\exp\left(\,-\, \frac{K}{3}\right) = \left( T+\overline{T} \right) \exp V $ \\[-4pt] \\[-4pt]
 $\left. W^\alpha\left(\frac{L}{S_0\overline{S}_0}\right) \, W_\alpha\left(\frac{L}{S_0\overline{S}_0}\right)\right|_F$ & & $W^\alpha(V)\, W_\alpha(V)$ \\[-4pt] \\[-4pt]
\hline\hline
\end{tabular}
%}
\caption{\small New--Minimal Dualities}
\label{table_3}
\end{table}

\subsection{$D$--term Inflation} \label{sec:dterm}
$R+R^2$ supergravity, $D$--term inflation \cite{fklp,fkr,ffs}, $\alpha$--attractor scenarios \cite{fklp,klr,klrc}, no-scale inflationary models \cite{eno}  and $\alpha$--scale models \cite{roestsca,linde_alpha} have a nice SU(1,1)/U(1) hyperbolic geometry with curvature $R_\alpha=-2/3\alpha$ for the inflaton superfield,  and (see Table \ref{table_1})
\beq
n_s \approx 1 - \frac{2}{N}\ ,
\qquad r = \frac{12\alpha}{N^2} \ .
\eeq{135}
An appealing and economical class of models allows to describe any potential of a single scalar field which is the square of a real function \cite{fklp},
\beq
 V(\varphi) \ = \ \frac{g^2}{2} \ P^2(\varphi) \ .
\eeq{136}
These are the $D$-term models, which describe the self-interactions of a massive vector multiplet whose scalar component is the inflaton. Up to an integration constant (the Fayet-Iliopoulos term), the potential is fixed by the geometry, since the K\"ahler metric is
\beq
ds^2 \ = \ ( d \varphi)^2 \ + \ \left( P^\prime(\varphi)\right)^2\, da^2 \ .
\eeq{14}
After gauging the field $a$ is absorbed by the vector, via $da+gA$, giving rise to a mass term $\frac{g^2}{2} \left(P^\prime(\varphi) \right)^2\, A_\mu^2$  via the BEH mechanism. The scalar potential is given in eq.~\eqref{136}
and the Starobinsky model corresponds to (Table \ref{table_3})
\beq
P(\varphi) \ = \ 1 \ - \ e^{\,-\,\sqrt{\frac{2}{3}} \, \varphi} \ .
\eeq{15}

In these models there is no superpotential, $V>0$, and only de Sitter plateaux are possible. At the end of inflation $\varphi=0$, $D=0$ and supersymmetry is recovered in Minkowski space.

\subsection{Other Models} \label{sec:othermodels}

Several examples exist with two chiral multiplets of the same sort, for which $F_S$ determines the dS plateau, $F_T = 0$, while at the end of inflation $F_S = F_T = 0$ and supersymmetry is recovered.

In a class of models ($\alpha$--attractors) the K\"ahler geometry remains SU(1,1)/U(1) for the inflaton subsector, as in the original $R+R^2$ theory, but the superpotential is modified \cite{klr} :
\beq
K \ = \ - \ 3\,\alpha \ \log \left( 1 \ - \ \frac{S\,\overline{S}\, + \, T \overline{T}}{3} \right), \quad W(S,T) \ = \ S\,\left( 3 \, - \, T^2 \right)^\frac{3\alpha-1}{2}f\left(\frac{T}{\sqrt{3}}\right)\ .
\eeq{18}
Along the inflationary trajectory the curvature reduces to $R(T)=-2/3\alpha$ and $V \ \sim \left| f \right|^2 \geq 0$.

In an alternative class of models the K\"ahler geometry is flat, with
\beq
K \ = \ \frac{1}{2} \left( T + \overline{T} \right)^2 \ + \ S\, \overline{S} \ , \quad W \ = \ S\, f(T) \ .
\eeq{19}
The inflaton originates from $Im T$, so that it does not enter $K$, which avoids a dangerous exponential factor from $e^K$ in the potential (and thus the so--called $\eta$--problem \cite{shift_eta}). This model belongs to a class where the K\"ahler potential has a shift symmetry \cite{klrube}. With a trivial K\"ahler geometry, along the inflationary trajectory the potential reduces to
\beq
V(\varphi) \ \sim \ \left| f(\varphi) \right|^2\ ,
\eeq{20}
so that the inflaton potential is fully encoded in the superpotential shape.
\begin{table}
\centering
$$ {\cal L} \ \sim \ \left. - \Phi \, S_0\, \overline{S}_0  \right|_D \ + \ \left. \left(W\, S_0^3 + {\rm h.c.} \right)\right|_F \ , \quad \Phi = \exp\left(\,-\, \frac{K}{3}\right)$$
\begin{tabular}{ccc}
\hline\hline
Higher Curvature & & Standard Supergravity \\
\hline\hline
 \\[-4pt]
%\hline
 $\Phi_H = 1 - \frac{1}{M^2}\ \frac{{\cal R}}{S_0}\ \frac{\overline{{\cal R}}}{\overline{S}_0} $ & & $\Phi_S = T + \overline{T} - S \, \overline{S} $ \\[-5pt] \\[-5pt]
 $W_H = W_0 + \xi \, \frac{{\cal R}}{S_0} + \sigma \, \frac{{\cal R}^2}{S_0^2}$ & & $W_s =  M T S + f S + W_0$ \\
 & & $\left(S^2 = 0 \ , \quad f = \xi - \frac{1}{2}\right)$ \\[-4pt]
\\[-4pt]
\hline \\[-4pt]
 $\Phi_H =  - \frac{1}{M^2}\ \frac{{\cal R}}{S_0}\ \frac{\overline{{\cal R}}}{\overline{S}_0} $ & & $\Phi_S = T + \overline{T} - S \, \overline{S} $ \\[-4pt] \\[-4pt]
 $W_H = \sigma \, \frac{{\cal R}^2}{S_0^2}$ & & $W_s =  M T S$ \\
 & & $\left(S^2 = 0 \right)$ \\[-4pt]
\\[-4pt]
 \hline \\[-4pt]
 $\Phi_H = 1 $ & & $\Phi_S = 1 - S \, \overline{S} $ \\[-3pt] \\[-3pt]
 $W_H = W_0 + \sigma\left(\frac{{\cal R}}{S_0} - \lambda \right)^2$ & & $W_s =  f S + W_0$  \\
 & & $\left(S^2 = 0 \ , \quad f = \lambda - 3 W_0 \right)$ \\[-4pt] \\[-4pt]
\hline\hline
\end{tabular}
\caption{\small Nilpotent old--Minimal Dualities}
\label{table_4}
\end{table}

\subsection{Nilpotent Inflation (sgoldstino--less models)}\label{sec:sgoldstinoless}

The problem with the models presented so far resides in the difficulty to obtain an exit from inflation with a supersymmetry breaking scale much lower than the de Sitter plateau scale (Hubble scale during inflation).

A way to solve this problem is to introduce a nilpotent sgoldstino multiplet $S$ \cite{quadratic} ($S^2=0$), so that the goldstino lacks a scalar partner. S is the Volkov-Akulov superfield. In this way the stabilization problem is overcome and a de Sitter plateau is obtained.

The first examples of cosmological models with a nilpotent sgoldstino multiplet was a generalization of the V-A-S supergravity of \cite{adfs}, with \cite{fkl}
\beq
W(S,T) \ = \ S \, f(T) \ , \qquad V \ = \ e^{K\left(T,\overline{T}\right)} \
K^{-1}_{S\bar{S}} \ \left| f(T) \right|^2 \ .
\eeq{21}
Models which incorporate separate scales of supersymmetry breaking during and at the exit of inflation have a trivial (flat) K\"ahler geometry
\beq
K(\Phi,S) \ = \ \frac{1}{2} \left( \Phi \ + \
\overline{\Phi} \right)^2 \ + \ S\, \overline{S}\ .
\eeq{22}
They in the supersymmetry breaking patterns during and after inflation:
\begin{itemize}
\item  in the first class of models \cite{kl14}
\beq
W(\Phi,S) \ = \ M^{\,2} \, S \left( 1 \ + \
g^2(\Phi) \right) \ + \ W_0 \ ,
\eeq{23}
where $g(\Phi)$ vanishes at $\Phi=0$ and the inflaton $\varphi$ is identified with its imaginary part.  Along the inflaton trajectory   $Re(\Phi)=0$                      is then
\beq
V \ = \ M^4 \, \left| g(\varphi)\right|^2 \left(2 \ +
\ \left| g(\varphi)\right|^2 \right) \ + \ V_0 \ ,
\quad V_0 = M^4 - 3\, W_0^2 \ .
\eeq{24}

Assuming $V_0 \simeq 0$, one finds
\beq
m_{\frac{3}{2}} = \frac{1}{\sqrt{3}}\ H \, \quad
E_{SB} = |F_S|^{\frac{1}{2}}=\sqrt{H M_P} > H
\eeq{25}
\beq
V \ = \ F_S\, F^S \ - \ 3\, W_0^{\,2} \ , F_\Phi=0 \
{\rm during \ inflation} \ (Re\Phi=0) \ ;
\eeq{26}
\item in the second class of models \cite{dz}
\beq
W(\Phi,S) \ = \ f(\Phi) \left( 1 \ + \ \sqrt{3}\, S \right)\ ,
\eeq{27}
which combines nilpotency and no-scale structure. Here:
\beq
\overline{f}(\Phi) \ = \
f(- \overline{\Phi}) \ , \ \  f^\prime(0)=0 \, \ f(0) \neq 0 \ .
\eeq{28}
The scalar potential is of no-scale type $\left(\Phi = \frac{1}{\sqrt{2}} \ (a + i\, \varphi) \right)$
\bea
&& F^S\,F_S \ = \ 3\, e^G \ = \ 3\, e^{a^2} \left| f(\Phi) \right|^{\,2} \ ,\nonumber\\
&& V(a,\varphi) \ = \ F^\Phi\,F_\Phi \ = \ e^{a^2} \left| f^\prime(\Phi) \, +\, a\, \sqrt{2}\, f(\Phi) \right|^{\,2} \ .
\eea{29}
$a$ is stabilized at 0 since $f$ is even in $a$. During inflation $a$ gets a mass O(H) without mass mixing with $\Phi$ and is rapidly driven to $a$=0.

The inflationary potential is
\beq
V(a=0,\varphi) \ = \ \left| f^\prime\left(
\frac{i \varphi}{\sqrt{2}}\right)\right|^{\,2} \ , \quad V(0,0)=0 \ .
\eeq{30}
These models lack the fine-tuning of the previous class ($V_0=0$).
It is interesting to compare the supersymmetry breaking patterns. Here $F_S$ never vanishes, and at the end of inflation
\beq
F^S\, F_S \ = \ 3 \ e^{G(0,0)} \ = \
3 \ m_\frac{3}{2}^2 \ .
\eeq{31}
More in detail,
\beq
\langle F^S \rangle_{\Phi=0} \ = \ \sqrt{3} \ \overline{f}(0) \ , \quad
m_{\frac{3}{2}} \ = \ \left| f(0) \right| \ ,
\eeq{32}
and the inflaton potential vanishes at the end of inflation. A choice that reproduces the Starobinsky potential is
\beq
f(\Phi) \ = \ \sigma \ - \ i\, \mu_1 \, \Phi \ + \ \mu_2 \
e^{\,i\,\frac{2}{\sqrt{3}}\, \Phi} \ .
\eeq{33}
Interestingly, $m_a,m_\frac{3}{2}$ depend on the integration constant $\sigma$ but $m_\varphi$ does not, since V does not depend on it.

\end{itemize}
\section{Higher-curvature and standard Supergravity duals}

In this section we describe dual higher--derivative \cite{stelle} formulations of some of the preceding models. Work in this direction started with the $R+R^2$ Starobinsky model, whose supersymmetric extension was derived in the late 80's \cite{cecotti,cfps} and was recently revived in \cite{eno,kl,fklp,fkr}, in view of new CMB data \cite{planck15}. Models dual to higher-derivative theories are subject to more restrictions than their bosonic counterparts or standard supergravity models. The three subsections are devoted to a brief description of $R+R^2$ supergravity, to a scale invariant $R^2$ Supergravity and to theories with a nilpotent curvature, whose duals describe non-linear realizations (in the form a Volkov--Akulov constrained superfield) coupled to supergravity.

\subsection{$R+R^2$ Supergravity} \label{sec:RR2}
The embedding of $R+R^2$ theories in Supergravity is sensitive to the choice of auxiliary fields, whose minimal sets result in the old minimal \cite{oldminimal} and new minimal \cite{newminimal} formulations of off--shell supergravity. There are thus two distinct classes of minimal models, depending on the choice of auxiliary fields.

Minimality resides in the fact that both sets only contain six bosonic degrees of freedom, which barely compensate the mismatch between the off--shell modes of a gravitino and a graviton, once gauge invariance is taken into account. In detail:
\begin{itemize}

\item Off-shell DOF's:   $g_{\mu\nu}$: 6(10-4) ,   $\psi_\mu$: 12(16-4)
$n_B= n_F$ off shell requires six extra bosons;
\item old minimal: 	$A_\mu$ , S, P   (6 DOF’s);
\item new minimal: 	 $A_\mu:$ 3(4-1) , $B_{\mu\nu}:$ 3(6-3), (6 DOF’s, due to gauge invariance)\ .
\end{itemize}
The $12_B$+ $12_F$ DOF's must fill massive multiplets \cite{stelle,fgvn}:
\beq
{\rm Weyl^2\ : \ } (2,2(3/2),1) \ , \quad R^2_{old} \ :
\ 2(1/2,0,0) \ , \quad R^2_{new} \ : \ (1,2(1/2),0) \ .
\eeq{16}
We do not consider the tensor excitations arising from Weyl${}^2$, which lead to a non--unitary theory \cite{stelle}.
After superconformal manipulations, these two theories are seen to be equivalent to standard supergravity coupled to matter (Tables \ref{table_2} and \ref{table_3}). The new minimal formulation gives $D$-term inflation as described before: the massive multiplet in eq.~\eqref{16} can be regarded as arising from a Stueckelberg mechanism where a chiral multiplet $T$ is eaten by a vector multiplet, which thus acquires a mass. The potential is then given in eq.~\eqref{136}, using \eqref{15}, and the de Sitter plateau is seen to arise from the Fayet--Iliopoulos term. On the other hand, the old minimal formulation gives $F$--term inflation with the two chiral superfields $T$ (inflaton multiplet) and $S$ (sgoldstino multiplet).

The $T$ submanifold, common to both formulations, is $SU(1,1)/U(1)$ with curvature $R = -2/3$, and the no-scale structure of the K\"ahler potential is responsible for the universal expression along the inflationary trajectory, where $F_S \neq 0$, $F_T=0$ imply that $S$ can be identified with the sgoldstino, and the potential reduces to
\beq
V \ = \ \frac{g^2}{2} \, M_{Pl}^4 \left(1 \ - \
e^{\,-\,\sqrt{\frac{2}{3}}\ \varphi} \right)^2 \ .
\eeq{17}

Note that in this class of models the contribution of $h\left(S,\overline{S}\right)$ in Table \ref{table_2} to the K\"ahler potential (see table \ref{table_2}) cannot be quadratic as originally in \cite{cecotti}, since otherwise instabilities show up \cite{kl} along the $S$ and $ImT$ directions.

We may now comment on other theories that are built solely in terms of gravitational curvatures, the first class of which was originally considered in ref.~\cite{ketov}. These ${\cal F}({\cal R})$ theories rest on a
chiral function ${\cal F}$ of the chiral multiplet ${\cal R}$: they do not give a de Sitter plateau, but have usually AdS vacua \cite{fkp} or an AdS plateau. The only exceptions are the $\alpha$--scale models, which however have instabilities \cite{roestsca}. All these theories do not lead to Starobinsky--like inflaton potentials \cite{fkvp}. The other class of theories gives $f(R)$ in components, and is related to theories already considered in the gravity literature \cite{capozziello}. Their supersymmetric completions
are different in the old \cite{cecotti} and new minimal \cite{fklp2} formulations. In the former, they are dual to theories with four chiral multiplets, some of which appear as ghosts \cite{cecotti,dlt}, at least in the simplest examples. On the other hand, in the new minimal setting higher curvature terms combine into Born--Infeld corrections to the original massive vector Lagrangian dual to the $R+R^2$ supergravity. These theories deserve further study, although some of their features were already analyzed in \cite{fklp2}.

\subsection{Scale--invariant $R^2$ models} \label{sec:R2_models}
Supergravity theories with unconstrained superfields also include $R^2$ duals, whose non--supersymmetric counterparts describe standard Einstein gravity coupled to a massless scalar field in de Sitter space (Table \ref{table_2}). These theories were recently resurrected in \cite{klt,akklr}. The $R^2$ higher curvature supergravity was recently obtained in both the old and new minimal formulations \cite{fkp}. In the old minimal formulation, the superspace Lagrangian is
\beq
\left.\alpha \, {\cal R} \, \overline{\cal R}\right|_D \ - \ \left(
\left. \beta\, {\cal R}^3 \ + \ {\rm h.c} \right) \right|_F\ ,
\eeq{34}
where
\beq
{\cal R} \ = \ \frac{\Sigma(\overline{S}_0)}{
{S}_0} \ \ (w=1,n=1) \ , \quad
{\overline{\cal D}}_{\dot{\alpha}} \, {\cal R} \ = \ 0
\eeq{35}
is the scalar curvature multiplet. Here $(w,n)$ denote the superconformal Weyl and chiral weights: note that, in general, $\Sigma(V)$  is superconformal provided $n_V=w_V-2$. The dual standard supergravity has
\beq
K \ = \ - \ 3\, \log\left(T+\overline{T} \ - \ \alpha\, S \, \overline{S} \right) \ , \quad \ W \ = \ T\, S \ - \ \beta\, S^3\ ,
\eeq{36}
where the K\"ahlerian manifold is SU(2,1)/U(2). Note the rigid scale invariance of the action under
\beq
T \ \rightarrow \ e^{\,2\,\lambda}\, T \ , \quad
S \ \rightarrow \ e^{\,\lambda}\, S \ , \quad S_0 \ \rightarrow
\ e^{\,-\,\lambda}\, S_0\ , \quad {\cal R} \ \rightarrow \ {\cal R} \ .
\eeq{37}
If $\alpha=0$ S is not dynamical, and integrating it out gives an SU(1,1) $\sigma$--model with
\beq
K \ = \ - \ 3\, \log\left(T+\overline{T} \right) \ , \ \ \ W \ = \ \frac{2\, T^{\frac{3}{2}}}{3\, \sqrt{3\, \beta}}\ .
\eeq{38}

In the new--minimal formulation the dual theory corresponds to the de--Higgsed phase
of the dual companion $R+R^2$ supergravity. The standard supergravity Lagrangian is then
\beq
{\cal L} \ = \ - \ \left. S_0\,\overline{S}_0 \, e^{gV}  \left(T+\overline{T}\right)\right|_D \ + \ \left. \frac{1}{4} \ W^\alpha(V)\, W_\alpha(V)\right|_F \ + \ {\rm h.c.} \ ,
\eeq{388}
and its bosonic terms read
\beq
e^{-1}\, {\cal L} \ = \ \frac{1}{2} \ R \ - \ \frac{1}{4} \ F^{\mu\nu}\,F_{\mu\nu} \ -\ \frac{3}{\left(T + \overline{T} \right)^2} \ \partial^\mu T\, \partial_\mu \overline{T} \ - \ \frac{1}{2}\ g^2 \ .
\eeq{389}
Note that the potential is just a positive cosmological constant, corresponding to a de Sitter maximally symmetric space. It is a slight modification of the model derived in \cite{fr76}, which does not include the $T$ multiplet.

\subsection{Nilpotent curvatures and sgoldstino-less Supergravity duals}

Higher--curvature supergravities can be classified via the nilpotency property of the chiral curvature ${\cal R}$, which gives rise to dual theories with nilpotent chiral superfields \cite{adfs}. The constraint
\beq
{\cal R}^2 \ = \ 0
\eeq{39}
in $R+R^2$ ($R$ is the bosonic scalar curvature) generates a dual theory where the inflaton chiral multiplet $T$ (scalaron) is coupled to the Volkov-Akulov multiplet $S$
($S^{\,2} = 0 \ , \ {\overline{\cal D}}_{\dot{\alpha}}\, S = 0$).
For this V-A-S Supergravity (Table \ref{table_4}) \cite{adfs}
\beq
K \ = \ - \ 3\, \log\left(T+\overline{T} \ - \ S \, \overline{S} \right) \ , \ \ \ W \ = \ g\, S\, T \ + \ f\, S \ + \ W_0 \ ,
\eeq{41}
and due to its no-scale structure the scalar potential is semi-positive definite
\beq
V \ = \ \frac{|g T \,+ \, f|^2}{3 \, ( T \, + \, {\overline T})^2} \ .
\eeq{42}
In terms of the canonically normalized field
\beq
T \ = \ e^{\,\phi{\sqrt \frac{2}{3}}} \ + \ i \, a \, \sqrt{\frac{2}{3}} \ , \qquad \left( \phi, a\right) \
{\in } \ \frac{SU(1,1)}{U(1)}
\eeq{43}
it becomes
\beq
V \ = \ \frac{g^{\,2}}{12} \left(1 \ - \ e^{\,-\,{\sqrt
\frac{2}{3}} \ \phi}\right)^2 \ + \ \frac{g^2}{18} \ e^{- 2\,\phi
{\sqrt \frac{2}{3}}} \ a^2 \ .
\eeq{44}
Here $a$ is an axion, which is much heavier than the inflaton during inflation
\beq
m_{\phi}^2 \ \simeq \ \frac{g^2}{9} \ e^{\,-\,2\, \phi_0 {\sqrt \frac{2}{3}}} \ << \
m_{a}^2 \ \equiv \ \frac{g^2}{9} \ .
\eeq{45}

There are then only two natural supersymmetric models with genuine single-scalar-field $\phi$ (inflaton) inflation: the new-minimal $R+R^2$ theory, where the inflaton has a massive vector as bosonic partner, and the V-A-S (sgoldstino-less) supergravity just described.

Another interesting example is the sgoldstino-less version of the ${\cal R} \overline{\cal R}$ theory described in Section \ref{sec:R2_models} (Table \ref{table_4}). This obtains imposing the same constraint as for the V-A-S supergravity,  ${\cal R}^2 \ = \ 0$  \cite{fps}, and is dual to the V-A-S supergravity with $f \ = \ W_0 \ = \ 0 $. The bosonic terms of the Lagrangian read
\begin{equation}
e^{-1}\, {\cal L} \ = \ \frac{R}{2} \ - \ \frac{1}{2}\ (\partial \phi)^2 \ - \ \frac{1}{2} \ e^{\,-\,2\, \phi \,{\sqrt \frac{2}{3}}}\
(\partial a)^2 \ - \ \frac{g^{\,2}}{12} \ - \ \frac{g^{\,2}}{18} \ e^{- 2\,\phi
{\sqrt \frac{2}{3}}} \ a^2 \ , \label{455}
\end{equation}
so that the potential
\beq
V \ = \ g^2 \ \frac{|T|^2}{3 ( T \,+ \,{\overline T})^2} \ = \ \frac{g^{\,2}}{12} \ + \ \frac{g^{\,2}}{18} \ e^{- 2\,\phi
{\sqrt \frac{2}{3}}} \ a^2
\eeq{46}
is positive definite and scale invariant.
In the maximally symmetric case, this model yields a de Sitter geometry with a positive vacuum energy
\beq
V\left(a=0\right) \ = \ \frac{g^{\,2}}{12}
 \ M_{Planck}^{\,4} \ .
\eeq{47}
This model also provides an approximation to the V-A-S model in the plateau region, where the potential becomes almost constant so that an approximate shift symmetry for $\phi$ emerges. This is precisely the scaling symmetry of the pure $R^2$ theory, which is present with both linear and non--linear realizations of the broken supersymmetry.

The de Sitter geometry plays a key role in the dynamics. Retracing the steps that led in \cite{adfs} to the dual higher--curvature form of the V-A-S model now yields
\begin{equation}
e^{-1}\,{\cal L} \ = \ \frac{3}{4 g^2} \, \left(R\,+ \, \frac{2}{3}\ A_m^2\right)^2 \  +  \ \frac{3}{g^2} \ ({\cal D} \cdot A)^2
\ . \label{475}
\end{equation}
This Lagrangian also enjoys an invariance under global scale transformations, whereby $A_m$ is inert but the metric transforms, since the new metric is related to the one in eq.~\eqref{455} by a Weyl rescaling involving $\phi$. Moreover, the proper scalar degree of freedom encoded in $D \cdot A$ emerges precisely around dS space, where $R = g^2/3$. This can be seen from the mixed term in eq.~\eqref{475}, which results in a linearized equation around dS space,
\beq
\partial_m \, ({\cal D} \cdot A) \ - \ \frac{g^2}{9} \ A_m \ = \ 0\ ,
\eeq{476}
with a mass term that is precisely the one carried by the axion field in eq.~\eqref{455}, as pertains to a dual formulation.

In contrast, the Volkov-Akulov model coupled to supergravity \emph{involves two parameters} and its \emph{vacuum energy has an arbitrary sign}. The pure V-A theory coupled to supergravity has indeed a superfield action determined by \cite{adfs}
\beq
K \ = \ 3\ S \, \overline{S} \ , \quad
W \ = \ f\, S \ + \ W_0 \ , \quad S^{\,2} \ = \ 0 \ ,
\eeq{48}
with a resulting cosmological constant \cite{deszum}
\beq
V \ = \ \frac{1}{3} \
\left|f\right|^2 \ - \ 3\ \left| W_0 \right|^2 \ .
\eeq{49}
Its component expression, including all fermionic terms, was recently worked out in \cite{va_components}. Note the invariance of \eqref{48} under $f \to - f$ combined with $S \to - S$. In the dual theory the corresponding transformations are $\lambda \to 6\, W_0 - \lambda$ and ${\cal R} \to 6\, W_0 - {\cal R}$.

The higher-curvature supergravity dual \cite{dfks,ignatiosr2,hasyam} is given by the standard (anti-de Sitter) supergravity Lagrangian augmented with the nilpotency constraint
\beq
\left(\frac{\cal R}{S_0} \ - \ \lambda\right)^2\ = \ 0\ .
\eeq{50}

This is equivalent to adding to the action the term
\beq
\left. \sigma
\left(\frac{\cal R}{S_0} \ - \ \lambda\right)^2 S_0^{\,3} \ + \ {\rm h.c.} \right|_F \ ,
\eeq{51}
where $\sigma$ is a chiral Lagrange multiplier.

A superfield Legendre transformation and the superspace identity \cite{cecotti}
\beq
(\Lambda + {\overline \Lambda}) S_0 \,
{\overline S}_0\bigg|_D \ = \ \Lambda \, {\cal R} \, S_0^2 \ + \ {\rm h.c.}\bigg|_F
\eeq{52}
which holds, up to a total derivative, for any chiral superfield, turn the action into the V-A superspace action coupled to standard supergravity with                               $f \, = \, \lambda \, - \, 3\, W_0$, and supersymmetry is broken whenever                                  $f  \, \neq \, 0$.
The higher-derivative formulation is peculiar in that the goldstino G is encoded in the Rarita--Schwinger field \cite{dfks}. At the linearized level around flat space
\beq
G \ = \ - \ \frac{3}{2\,\lambda} \left( \gamma^{\mu\nu}\,
\partial_\mu\, \psi_\nu \ - \ \frac{\lambda}{2} \ \gamma^\mu\,
\psi_\mu \right) \ , \ \ \delta\, G \ = \ \frac{\lambda}{2} \
\epsilon \ .
\eeq{53}
The linearized equation of motion for the gravitino is
\beq
\gamma^{\mu\nu\rho}\, \partial_\nu\, \psi_\rho \ - \ \frac{\lambda}{6}\ \gamma^{\mu\nu}\, \psi_\nu \ - \ \frac{1}{3} \left(\gamma^{\mu\nu}\, \partial_\nu\ -
\frac{\lambda}{2} \ \gamma^\mu\right) G \ = \ 0 \ ,
\eeq{54}
and is gauge invariant under
\beq
\delta\, \psi_{\mu} \ = \
\partial_\mu \, \epsilon \ + \ \frac{{\lambda}}{6} \ \gamma_\mu \, \epsilon \ .
\eeq{55}
The $\gamma$--trace and the divergence of the equation of motion both give
\beq
\gamma^{\mu\nu}\, \partial_\mu\, \psi_\nu \ - \
\gamma^\mu\, \partial_\mu \ G \ = 0 \ .
\eeq{56}
Gauging away the Goldstino G one recovers the standard formulation of a massive gravitino.

\section{Sgoldstino-less Models \emph{vs} String Theory: Climbing Scalars}

In String Theory \cite{stringtheory} exponential potentials, with exponents implied by the Polyakov series, emerge due to uncanceled tensions $T$ from branes and orientifolds in ``brane supersymmetry breaking'' \cite{bsb}, a phenomenon that presents itself in orientifold vacua \cite{orientifolds} containing $O_+$--planes (with $T>0,Q>0$), and anti D-branes (with $T>0,Q<0$). The key role of non--linear realizations of supersymmetry for these systems was recognized long ago by Dudas and Mourad in \cite{dm}. Their four--dimensional analogs afford simple realizations within the class of sgoldstino-less models discussed in Section \ref{sec:sgoldstinoless} \cite{fkl,wrase}.

Exponential potentials result in interesting cosmological solutions that have been explored since the 1980's \cite{lm,exponential}, but a key feature went unnoticed for a while. This is the \emph{climbing behavior} in the presence of \emph{``(over)critical''} exponentials \cite{dks}, which we would like to explain. Here we shall sketch some of the main properties of these systems and their links with Section \ref{sec:sgoldstinoless}, leaving a more detailed review of their applications to the CMB for \cite{as_erice}.

Let us begin by recalling that a four--dimensional scalar field $\phi$ minimally coupled to gravity evolves in a spatially flat FLRW Universe
\beq
ds^2 \, = \, -\, dt^2 \, + \, e^{\,2\,A(t)} \,d{\bf x}\cdot d{\bf x} \ ,
\eeq{551}
according to
\beq
\ddot{\phi} \,+ \,  3 \, \dot{\phi} \, \sqrt{\frac{1}{6} \ \dot{\phi}^{\,2} \, + \, \frac{1}{3} \ V(\phi)} \, + \, V^{\,\prime} \nonumber  \ = \ 0 \ .
\eeq{561}
The competing effects of the driving force exerted by the potential and of the damping term can result in slow-roll. It is well known that this balance can be quantified via the slow--roll parameter $\epsilon$, but here we would like to describe a less familiar choice for the time coordinate, available in regions where the potential $V$ does not vanish, which is quite instructive.

In $D$ dimensions, and with an unconventional time variable $\tau$ tuned to the potential $V$ according to
\beq
ds^{\,2} \ = \ e^{\,2\,{\cal B}(t)} \, dt^2 \ - \ e^{\,\frac{2\,{\cal A}(t)}{d-1}}\
d{\bf x} \,\cdot \, d{\bf x} \ ,  \qquad V \, e^{\,2\,{\cal B}} \ = \ V_0 \, \qquad \tau \ = \ t\, \sqrt{\frac{D-1}{D-2}} \ ,
\eeq{57}
the equation of motion for an inflaton $\phi$ rescaled according to
\beq
\varphi \ = \ \phi\, \sqrt{\frac{D-1}{D-2}}
\eeq{58}
turns into
\beq
\ddot{\varphi} \ + \ \dot{\varphi} \sqrt{1 \ + \ { \dot{\varphi}^{\,2}}} \ +
\ \frac{V_{\varphi}}{2\, V} \, \left( 1 \ + \ { \dot{\varphi}^{\,2}} \right) \ = \ 0 \ .
\eeq{59}

There is a striking analogy between \eqref{59} and the equation for a Newtonian particle moving with velocity $v$ in a one--dimensional viscous fluid under the influence of an external driving force $f$,
\beq
\dot{v} \ + \ \frac{\beta}{m} \ {v} \ = \ \frac{f}{m} \ .
\eeq{60}
For $\left|\dot{\varphi}\right|<<1$ the two equations coincide, up to the identification $f \longleftrightarrow  - \, \frac{V_{\varphi}}{2\, V}$ .

The analogy makes manifest, for instance, why Linde's chaotic inflation \cite{inflation} occurs, in the quadratic potential $V = V_0 \left(\varphi/\varphi_0\right)^2$, when the field is far away from the origin. On the other hand, for an exponential potential
\beq
V = V_0 \ e^{\,-\,2\gamma\,\varphi}\ ,
\eeq{62}
where we can take $\gamma>0$ for definiteness, the driving term is simply $-\gamma$, a constant, so that inflation can either occur everywhere or nowhere, depending on its actual value. A closer look would reveal that, in $D$ dimensions, the upper bound for $\gamma$ to have inflation is $\frac{1}{\sqrt{D-1}}$ \cite{dks}.

A uniform motion at the limiting speed $v_{lim}=\frac{f}{\beta}$ is an exact solution for the Newtonian system \eqref{60} with $f$ constant and $\beta \neq 0$, and moreover \emph{all} solutions approach the limiting speed asymptotically. A sharp transition occurs however as $\beta \to 0$, since the motion becomes uniformly accelerated. Eq.~\eqref{59} exhibits a similar transition at $\gamma=1$, and indeed for $0<\gamma<1$ a limiting speed exists, $\dot{\varphi}_{lim} \ = \ \frac{\gamma}{\sqrt{1-\gamma^2}}$\ .

Moreover, a uniform motion at $\dot{\varphi}_{lim}$ solves exactly also eq.~\eqref{59}: it is the form taken by the Lucchin--Matarrese attractor \cite{lm} in this gauge. A sharp change in the nature of the solutions occurs, however, at $\gamma=1$ \cite{dks}: for any $0<\gamma<1$ the system \eqref{59} admits two independent types of solutions, whereby the scalar emerges from the initial singularity \emph{climbing} or \emph{descending} along the exponential potential, while \emph{for $\gamma\geq 1$ the descending solution disappears}.
This can be seen analyzing the limiting behavior in the potential \eqref{62} close to the initial singularity, where eq.~\eqref{59} reduces to
\beq
\ddot{\varphi} \ + \ \dot{\varphi} \left|\dot{\varphi}\right| \ -
\ \gamma \, \dot{\varphi}^{\,2} \ = \ 0 \ .
\eeq{592}
The corresponding solutions behave as $\dot{\varphi} \ \sim \ C/\tau$, with
\beq
\left|C\right| \ = \ \frac{1}{1 \ - \ \gamma\, {\rm sign}(C)} \ ,
\eeq{593}
and if $\gamma >1$ this condition implies that $C<0$, so that the scalar is bound to start out while climbing up the exponential potential. On the other hand if $\gamma<1$ both signs are possible for $C$, and a descending solution also exists. The case $\gamma=1$ is solved exactly by
\beq
\dot{\varphi} \ = \  -  \ \frac{1}{2\, \tau} \ + \ \frac{\tau}{2} \ ,
\eeq{594}
which also describes a climbing scalar.

Intriguingly, the Polyakov expansion implies that the Sugimoto model of \cite{bsb} sits precisely at $\gamma=1$, which we shall refer to as the ``critical'' value, and it is thus solved exactly by eq.~\eqref{594}. To reiterate, for $\gamma \geq 1$ the system can be termed a ``climbing scalar'', since the scalar climbs up the exponential potential as it emerges from the initial singularity, before reverting its motion and starting to descend. In this dynamics the string coupling is bounded from above, so that this picture is naturally protected against string loops (although not against $\alpha^\prime$ corrections), and \emph{similar considerations apply whenever the potential is dominated asymptotically by an (over)critical exponential} \cite{dks}.
In lower dimensions $\varphi$ is accompanied by the breathing mode of the internal space. Amusingly, however, for all $D$ one of their orthogonal combinations maintains a critical exponential potential \cite{as13}. If the other were somehow stabilized, a fate for which, unfortunately, no clues are available at this stage from string corrections, a climbing scalar would be inherited in four dimensions, where the resulting systems fall nicely into the pattern of Section \ref{sec:sgoldstinoless}.

When combined with other terms capable of sustaining an inflationary phase, a ``critical'' exponential can lead to an interesting behavior, whereby inflation is preceded by a climbing phase that ends in a bounce. The scalar, momentarily in slow-roll as it reverts it motion, can give rise to a new class of features in primordial power spectra, pre--inflationary peaks lying well apart from the almost scale invariant Chibisov--Mukhanov profile. The simplest models of this type rest on the two--exponential potentials
\beq
V \ = \ V_0 \left( e^{\,-\,2\,\varphi} \ + \ e^{\,-\,2\,\gamma\, \varphi} \right)\ ,
\eeq{64}
and with $\gamma = \frac{1}{12}$, a value that will recur shortly, they yield a spectral index $n_s \approx 0.96$, which is close to the currently preferred value $n_s \approx 0.97$ of \cite{planck2015}. If the CMB were confronting us with signs of the onset of inflation, angular power spectra determined by potentials close to \eqref{64} could follow nicely the first few CMB multipoles of \cite{planck2015}, with a low quadrupole, some features around $\ell=5$ and a dip around $\ell=20$ \cite{ks}. Provided, of course, the 7--8 $e$--folds accessible to us via the observed CMB were granting us access to the transition region. Or, if you will, if we were observing not only imprints of a well developed slow-roll, but also of its onset.

Curiously, a $p$--brane coupling to the dilaton in the string frame as $e^{-\,\alpha \phi}$, under assumptions similar to those leading to a climbing scalar in lower dimensions, would result in exponents $\gamma$ that are multiples of $\frac{1}{12}$ \cite{as13},
\beq
\gamma \ = \ \frac{1}{12} \left( p \ + \ 9 \ - \ 6\, \alpha \right) \ .
\eeq{65}
This numerology points once more to the interest for these types of potentials, but two--exponential models are clearly not realistic. However, the basic lesson also applies to more realistic potentials (in four dimensions $\sqrt{6}\,\phi=2 \varphi$)
\beq
V(\phi) \ \sim \ e^{\,-\,\sqrt{6} \,\phi} \ + \ v(\phi) \ ,
\eeq{66}
which include a ``critical'' exponentials and where can sustain inflation and allow fro a graceful exit: the scalar $\varphi$ continues to exhibit a climbing behavior and resulting power spectra can be very similar. These models can nicely account for the first few multipoles, but unfortunately they tend to approach too slowly the attractor behavior. Or, if you will, they tend to modify the CMB angular power spectrum well beyond the first 20-30 multipoles, where an anomalous behavior with respect to the reference $\Lambda$CDM model manifests itself.

Actually, in detailed comparisons with the CMB one has little freedom to depart from $\Lambda$CDM without spoiling its remarkable determination of the cosmological parameters. A minimal well motivated choice is \cite{cmbcompare}
\beq
k^{\,n_s-1} \ \longrightarrow \ \frac{k^{\,3}}{\left(k^2 \ + \ \Delta^2\right)^{2-\frac{n_s}{2}}} \ ,
\eeq{666}
which can model the typical infrared cut associated to a fast inflaton. One can thus determine, with some subtleties related to the choice (or art) of Galactic masking, a cosmological scale of $\Delta \sim 2800$ Mpc related to quadrupole damping, to about 99\% confidence level. The evidence is stronger than in previous analyses carried out with standard Galactic masks \cite{standardpeak}, but the value of $\Delta$ is similar. It can translate into pre--inflationary energy scales ${\cal O}(10^{14})\ GeV$ \cite{cmbcompare} with a inflationary phase of about 65 $e$--folds of inflation.

We would like to conclude explaining how the potentials \eqref{66} can be simply embedded in four--dimensional supergravity resorting to a nilpotent scalar multiplet $S$, as in Section \ref{sec:sgoldstinoless}. To this end, one can let
\beq
K \ = \ - \ 3 \, \log\left(T+\overline{T}\right) \ + \ h\!\left(T,\overline{T}\right)\, {S \, \overline{S}} \ , \quad W \ = \ W_0 \ + \ f \ S \ .
\eeq{67}
and
\beq
h^{-1}\!\left(T,\overline{T}\right) \ = \ 1 \ + \ {\left(\frac{T \ + \ \overline{T}}{2} \right)^3}\ {v\!\left({T,\overline{T}}\right)} \ ,
\eeq{68}
which generalizes the choice in \cite{fkl,wrase}.
``Critical'' potential wells are special, and the reader is referred to \cite{dconde} for some amusing details.
\vskip 12pt

\noindent{\large \bf Acknowledgements}\\ \noindent
We would like to thank I.~Antoniadis, E.~Dudas, A.~Gruppuso, R.~Kallosh, A.~Kehagias, N.~Kitazawa, A.~Linde, N.~Mandolesi, P.~Natoli, M.~Porrati and A.~Van Proeyen for recent collaborations on related issues. This work was supported in part by INFN-CSN4 (I.S. GSS and Stefi). A.~S. was also supported in part by Scuola Normale, and would like to thank the CERN Th-Ph Department for the kind hospitality.
%
%%%%%%%%%%%%%%%%%%%%%%%%%%%%%%%%%%%%%%%%%%%%%%%%%%%%%%%

%%%%%%%%%%%%%%%%%%%%%%%%%%%%%%%%%%%%%%%%%%%%
\end{document}